\documentclass[preprint,prb,superscriptaddress,showpacs,preprintnumbers]{revtex4}
\usepackage{amsmath,dcolumn,graphicx,ulem}

\begin{document}

\title{Relationship between macroscopic physical properties and local distortions of low doping La$_{1-x}$Ca$_{x}$MnO$_3$: an EXAFS study}

\author{Y. Jiang} \affiliation{Physics Department, University of California,
Santa Cruz, California 95064, USA}
\author{F. Bridges} \affiliation{Physics Department, University of California,
Santa Cruz, California 95064, USA}
\author{L. Downward} \affiliation{Physics Department, University of California,
Santa Cruz, California 95064, USA}
\author{J. J. Neumeier} \affiliation{Department of Physics, Montana State University, Bozeman, Montana 59717, USA}

\date{\today}

\begin{abstract} 

A temperature-dependent EXAFS investigation of La$_{1-x}$Ca$_{x}$MnO$_3$ is
presented for the concentration range that spans the ferromagnetic-insulator
(FMI) to ferromagnetic-metal (FMM) transition region, x = 0.16, 0.18, 0.20, and
0.22; the titrated hole concentrations are slightly higher  y = 0.2, 0.22,
0.24, and 0.25 respectively.  For this range of Ca concentrations the samples
are insulating for x = 0.16-0.2 and show a metal/insulator transition for x =
0.22.  All samples are ferromagnetic although the saturation magnetization for
the 16\% Ca sample is only $\sim$ 70\% of the expected value at 0.4T. This raises a
question as to the nature of the ferromagnetic (FM) coupling mechanism in such
insulating samples.  We find that the FMI samples have similar correlations
between changes in the local Mn-O distortions and the magnetization as observed
previously for the colossal magnetoresistance (CMR) samples (0.2 $\leq$x$\leq$ 0.5)
- except that the FMI samples never become fully magnetized. The data show that
  there are at least two distinct types of distortions. The initial distortions
removed as the insulating sample becomes magnetized are small and provides
direct evidence that roughly 50\% of the Mn sites (associated with the hole
charge carriers) have a small distortion/site and are magnetized first. The
large Mn-O distortions that remain at low T are attributed to a small fraction
($<$ 30\%) of fully Jahn-Teller-distorted Mn sites that are either unmagnetized
or antiferromagnetically ordered.  Thus the insulating samples are very similar
to the behavior of the CMR samples up to the point at which the M/I transition
occurs for the CMR materials. The lack of metallic conductivity for x$ \leq$ 0.2,
when 50\% or more of the sample is magnetic, implies that there must be
preferred magnetized Mn sites (that involve holes) and that such sites do not
percolate at these concentrations.

\end{abstract}
                                                                                                                                                               
\pacs{71.38.-k, 61.10.Ht, 71.27.+a, 75.47.Lx}

\maketitle


\section{Introduction} 

LaMnO$_3$ is an antiferromagnetic insulator and has a large Jahn-Teller (J-T)
distortion of the MnO$_6$ octahedron. When it's doped with Ca on the La sites
(i.e. La$_{1-x}$Ca$_{x}$MnO$_3$ (LCMO)), holes are introduced into the Mn e$_g$
band and this leads to novel transport and magnetic properties including
colossal magnetoresistance (CMR) and ferromagnetism for samples with Ca
concentrations, x, roughly in the range
20-50\%.\cite{Schiffer95,Imada98,Cheong00b} In this concentration range,
these pseudo-cubic manganites are paramagnetic insulators (semiconductors) at
high T, with a significant distortion of the Mn-O octahedron (for some sites)
that is associated with Jahn-Teller distortions and polarons.  Upon lowering
the temperature below the ferromagnetic transition temperature T$_c$, 
the Mn spins begin to align and thus the holes (electrons) can hop more rapidly between
Mn atoms without a spin flip via the intervening O atom; this enhances the
ferromagnetic (FM) coupling between Mn spins and is referred to as the double
exchange (DE) interaction.\cite{Zener51,Anderson55,deGennes60} To explain the
large magnetoresistance a polaron-like lattice distortion must also be
present.\cite{Millis95,Millis96,Millis96b,Millis96c,Roder96}
A consequence of very fast hopping of holes between neighboring
Mn sites is that the broadening, $\sigma$,  of the Mn-O pair distribution
function (PDF) is greatly reduced when the sample becomes magnetic, thus
$\sigma^2$(T) decreases rapidly below the ferromagnetic transition temperature
T$_c$.  Finally, at low T, LCMO in the CMR regime is a ferromagnetic metal with
very little static distortion of the Mn-O bonds.\cite{Booth98a,Booth98b,Downward2005}

For LCMO in this concentration range (x $\sim$ 0.2 - 0.5),  Downward {\it et
al.}\cite{Downward2005} have shown that there is a strong correlation between
the local Mn-O distortions removed, D = $\Delta (\sigma^2)$, and the
magnetization, as T is lowered through and below T$_c$.  By plotting D (for the
Mn-O PDF) versus the sample magnetization,
Downward {\it et al.}\cite{Downward2005} found that D increases
slowly with M until M/M$_0$ $\sim$ 2x (M$_0$ is the saturation magnetization at
low T); for larger M, D increases more rapidly. The low
initial slope in such a plot is direct experimental evidence of sites with a
low distortion/site, while the large slope ($\sim $ 4 times larger) at high M
indicates that the remaining small fraction of Mn sites have a large distortion/site.
Thus there are at least two different types of distortion present,
which have only been distinguished (so far) by correlating with magnetization.
An important question is - can one of the regimes be made more dominant by
appropriate doping, so that this effect is more directly observable? We will
address that issue in this paper.

Because the fraction of low distortion sites is roughly 2x, Downward {\it et
al.}\cite{Downward2005} proposed a two-site polaron model (called a dimeron) in
which a hole is partially delocalized over two Mn sites [initially one Mn site
would correspond to an e$_g$ electron site (Mn$^{+3}$) while the other would
correspond to a hole site (Mn$^{+4}$)].  
The magnetization, M, first develops via the aggregation of these
low-distortion dimeron sites (or multiples of such pairs).  The argument for a
lower distortion of the dimeron is that when the hole (or electron) is
partially delocalized -- a dynamic effect, the average charge per site is 
reduced to 3.5 (from 4 electrons on an isolated Mn$^{+3}$ site) and
J-T energies are reduced.  

In this model for M/M$_0$ $>$ 2x, further increases in M (induced by either
lowering T or increasing B) force the remaining J-T-distorted Mn sites, each
containing one e$_g$ electron, to become magnetized. If the hole quasiparticles
now spend time on such sites, a large J-T distortion per Mn site is removed as
observed for CMR samples.\cite{Downward2005} In fact, {\it most} of the
distortion removed in the FM state for CMR samples occurs {\it after} the
sample is more than 50\% magnetized.
Consequently, if the sample never became fully magnetized, e.g. for the lower
Ca concentrations considered here, the remaining net Mn-O distortion at low T
should be much larger than has been observed for the CMR regime.  Note that in
the partially magnetized state at low M there will be three types of Mn sites -
i) undistorted sites in the developing magnetic regions, ii) low distortion
sites corresponding to the remaining dimerons (polarons) that are not yet part
of a magnetic cluster, and iii) highly distorted Mn$^{+3}$ sites.

In considering how double exchange leads to both ferromagnetism and a
disappearance of local Mn-O distortions, an important question to ask is how
many Mn sites can one hole keep undistorted by hopping rapidly between them in
the ferromagnetic metallic phase?  The answer appears to be about four. For the
25\% Ca samples, there are four sites/hole, (for higher Ca concentrations fewer
sites/hole).  The material is ferromagnetic and there is little excess
distortion at low T for x between 25 and 40\%\cite{Booth98b,Downward2005} - the
Mn-O distortion at low T approaches the zero-point-motion value comparable to
that observed in CaMnO$_3$. For lower Ca concentrations, there is an increasing
amount of distortion at low T and at x = 0.21 the sample does not become fully
magnetic even at moderate fields (0.4T).\cite{Downward2005} Thus in this case
the holes do not get to every site often enough to maintain all the sites at a
low distortion.  

A related point is that dynamics plays an important role; if the holes hop
fairly rapidly above T$_c$ - at least among a few sites - then there will be no
Mn$^{+4}$ sites in the sample over the time scale of most experiments.  Thus
characteristics of a particular site becomes a time average of a hole site and
an electron site as holes hop on and off that particular site. This will depend
on the time scale of the measurements and the hole hopping rate; the latter
exceeds optical phonon frequencies in the FMM regime.  When the hole is present
the site becomes less distorted, whereas when an e$_g$ electron is present it
will become more distorted; consequently, rapid hopping at the local level will
produce a lower net distortion per site, at least for some sites. 

At lower Ca concentrations it is well known that the material becomes a
ferromagnetic insulator (semiconductor) at low T.\cite{Schiffer95} An important
detail is that these samples are never fully magnetized in moderate fields at
low T, and  the unmagnetized fraction increases slowly as the Ca concentration
decreases  below $\sim$20\% Ca. Since LCMO is a soft ferromagnet for higher Ca
concentrations it is fully magnetized at low T for moderate fields
(0.4T)\cite{Alonso00,Downward2005} and as we will show, for all the samples
that have a M/I transition. At lower Ca concentrations (x $\sim$ 0.1) a canted
AFM state exists\cite{Biotteau01,Pissas05} but disappears near x =
0.125.\cite{Biotteau01} However that does not preclude tiny AFM coupled
clusters a few nm in size whose size diminishes as x approaches 0.22. Thus the
nature of the magnetic structure is not well understood in this concentration
range.

At first ferromagnetism in an insulator appears to be in contradiction to a DE
model,\cite{Dai00,Papavassiliou01,VanAken03} because for double exchange, the
Mn spins are ferromagnetically coupled via fast hopping of the holes (or e$_g$
electrons) between Mn sites in the metallic regime. Some have suggested a
mixture of (DE) ferromagnetic metallic, and ferromagnetic insulating (distorted
and non-conducting) phases,\cite{Papavassiliou00,Papavassiliou01,Algarabel03}
but that requires invoking a new FM coupling mechanism (FM superexchange) for
the FMI phase (for only a tiny change in Ca concentration), although the
superexchange mechanism is normally AFM in LaMnO$_3$.  That may well be
necessary at low concentrations (x = 0.1) where there are not enough holes to
couple all the Mn sites via DE (the ratio is now 10 Mn sites/hole). Interestingly, the
x = 0.1 sample can be made completely ferromagnetic but non-conducting in a
12T field;\cite{Algarabel03} however in that case there should be a significant
fraction of the sample that is magnetized and yet retains a large J-T distortion
since the latter is only removed if the hole quasiparticles are hopping rapidly
over all the sites. One alternative explanation for a low magnetization at
moderate fields for concentrations near 16-20\% is that FM (conducting) domains
do form but are separated by insulating, non-ferromagnetic regions (possibly
tiny antiferromagnetically coupled clusters or a frustrated spin-glass-like configuration
as a result of competing FM and AFM magnetic interactions for some sites).
In either case, for x near 20\% a small fraction of the sample appears to block
the metallic conductivity and appears to be due to the intrinsic inhomogeneity
of the sample at the unit cell level.  


Several other papers also speak to this inhomogeneity. Electron magnetic
resonance (EMR) studies\cite{Markovich02,Shames03} indicate multiple phases in
a sample close to the concentration-driven MI transition.  For x = 0.18 and 0.2
they observe at least three resonance lines\cite{Shames03} which they attribute
to different phases just below T$_c$ - a ferromagnetic metallic phase and two
distinct ferromagnetic insulating phases.  Other investigations at low Ca
concentrations suggest that Mn$^{+4}$ hole sites may be localized.  Alonso {\it
et al.}\cite{Alonso00} argue that in such samples the conductive properties
depend on the origin of the holes - from metal atom vacancies or from Ca
dopants; the samples need Ca dopants to be conducting at higher concentrations.
They propose that localized Mn$^{+4}$ hole sites form close to the Ca dopants
which leads to unconnected magnetic/metallic clusters at low Ca concentrations;
they also suggest that perhaps the J-T distortions of Mn$^{+3}$ sites near the
Ca$^{+2}$ dopants are smaller than near La$^{+3}$, but they do not provide a
well defined model.  Algarabel {\it et al.}\cite{Algarabel03} also suggest that
holes are likely found close to the Ca sites in low concentration samples.
Finally, in zero-field NMR experiments using low Ca concentration samples, at
least 3 distinct regions with different hyperfine fields at the Mn nuclei have
been observed, but only at very low T when the magnetization has reached it's
maximum value;\cite{Papavassiliou00} they do not report results in the
temperature regime (100-200K) over which most of the magnetization develops.
They argue that a significant fraction of (localized) Mn$^{+4}$ are FM coupled.
This requires a different FM coupling mechanism to produce insulating FM
Mn$^{+4}$ sites and essentially no hopping of such holes.  

Finally it should be noted that diffraction studies\cite{Biotteau01,Pissas05}
suggest a transition for low Ca concentrations from a high temperature
pseudo-cubic to a low temperature, lower symmetry phase, with the transition
between 200-400K. Biotteau {\it et al.}\cite{Biotteau01} interpreted this
transition as an evolution from a dynamic Jahn-Teller effect (high T) to a
cooperative static J-T distortion (Space group {\it Pbnm}) at lower T.  A more
recent paper\cite{Pissas05} describes this transition as a symmetry change at
T$_{J-T}$ from  $P$2$_1/c$ at low T to orthorhombic ({\it Pnma}) at high T.
For example, for the x = 0.175 sample, Pissas {\it et al.}\cite{Pissas05}  show
that the separation of the Mn2-O bond lengths is a maximum near 200K for 
$P$2$_1/c$; the distribution of bond lengths narrows slightly at low T but
collapses to essentially one bond length at 300K.  This is in contrast  to the
EXAFS results we present here.  EXAFS is a very fast probe (10$^{-15}$ sec) and
can see similar J-T distortions in either the dynamic or static regime. We
discuss these and other results mentioned above, in the discussion section.


So far, there have been few local structure studies done for lower Ca
concentration samples, though we have previously shown \cite{Booth98b} that the
distortion removed as the sample becomes magnetized, is small for a low
concentration, x $\sim$ 0.12.  However, for samples at the boundary between the
ferromagnetic insulator region (Ca concentration below 20\%) and the CMR region
(Ca concentration $\sim$ 20-50\%), no local structure experiments have been
done to explore the relationship between distortions removed and the sample
magnetization.   Here, we present a detailed EXAFS study through the
concentration driven metal/insulator transition. We find similar
behavior as for LCMO with higher Ca concentrations, but the total distortion
removed in the FM insulating state at low T is greatly reduced.  Specifically,
in a comparison of 4 different samples [Ca concentrations 0.16 - 0.22; effective
hole concentrations y = 0.2 - 0.25] we find that the total magnitude of the
local Mn-O distortions removed in the FM state, decreases rapidly (by more than
a factor of 2) when the metal/insulator transition disappears in the
resistivity data.  For the insulating samples, the small overall change in
$\sigma^2$(T) as the sample becomes magnetized is direct evidence that a large
number of Mn sites (at least 50\%) have a low distortion per site at T$_c$.

This paper is organized as follows: a brief description of the sample preparation
and the EXAFS techniques are presented in Sec. II, the magnetization and
resistivity data are presented in Sec.  III,  and then the EXAFS data and
analysis are discussed in Sec. IV.  A comparison with the model of Downward
{\it et al.}\cite{Downward2005} and with other results is provided in Sec. V.

\section{Experimental details and the EXAFS technique}
\label{exp}

Transmission EXAFS Mn K-edge data were collected over a wide temperature range
{(3-550K)} on powdered samples at the Stanford Synchrotron Radiation Laboratory
(SSRL). A cryostat was used to collect the low temperature (3-300K) data at
beamlines 10-2 and 2-3 using both Si(111) and Si(220) monochromators, and an
oven was used for the high temperature (300-550K) data at beamline 2-3 using a
Si(111) monochromator. To reduce the harmonic content in the X-ray beam, we
detuned the monochromator crystals 50\%, and also used a harmonic rejection
mirror for the Si(220) monochromator on 10-2.  The samples are oriented 90
degree to the x-ray beams. The energy resolution ($\delta$E) is 1.0 eV for the
Si(111) monochromator and 0.44 eV for the Si(220) monochromator.

Samples of La$_{1-x}$Ca$_x$MnO$_3$ with x = 0.16, 0.18, 0.20, and 0.22 were
made by weight appropriate amounts of 99.99\% purity or better La$_2$O$_3$,
CaCO$_3$, and MnO$_2$, mixing with a mortar and pestle for 10 min, and reacting
for 3 h at 1150$^{\circ}$C in air in an alumina crucible. The samples were
removed from the furnace, reground for 10 min, and reacted for 16 h at
1250$^{\circ}$C. This last step was repeated 5 times with reaction temperatures
of 1300$^{\circ}$C, 1350$^{\circ}$C, 1375$^{\circ}$C (twice) and
1400$^{\circ}$C in air. The samples were then reground for 10 min, pressed into
pellets, and reacted for 16 h at 1400$^{\circ}$C.  The as-made samples, x =
0.18-0.22, showed a metal/insulator transition and had excess oxygen (i.e. a
few metal atom vacancies). To lower the O concentration, pressed pellets of the
samples were placed in a flow of Ar gas for 12 h at 1250$^{\circ}$C.
Iodometric titration was used to determine the final average Mn valence which
was found to be 3.20(1), 3.22(1), 3.23(1) and 3.25(1) for x = 0.16, 0.18, 0.20,
and 0.22, respectively. The excess O is about 0.015 (i.e. O$_{3.015}$), or in
terms of metal site vacancies, the composition would be approximately
(La$_{1-x}$Ca$_x$)$_{0.994}$Mn$_{0.994}$O$_3$. Thus the non-stoichiometry is
very small.  Powder x-ray diffraction revealed the single phase nature of the
samples. 

Magnetization versus T was measured in a magnetic field of 0.4T for most
samples (0.2T for the as made 0.16 and 0.18 samples); M vs B data were also
collected at low T. The electrical resistivity was measured on small bars cut
from the pressed pellet (1mm X 1mm X 6mm) using a four-probe dc technique. 



To prepare EXAFS samples, the pressed pellets were ground in a mortar and
pestle, passed through a 400-mesh sieve, and then brushed onto scotch tape for
the cryostat measurement, or onto kapton tape for the oven measurement. The
tape preferentially holds the smaller grains ($\le5 \mu m$) in a thin layer;
two double layers of tape were used for these EXAFS measurement.

Fits of the EXAFS data were carried out to the real and imaginary functions in
r-space (the Fourier Transform of k$\chi(k)$), using the EXAFS equation for 
k$\chi(k)$ which is given by:

\begin{eqnarray}
k\chi(k) &= & \sum_{i} k\chi_i(k) \nonumber \\
& = &Im \sum_{i} A_i
\int_{0}^{\infty} F_i(k,r)\frac{g_i(r_{0i},r)e^{i(2kr + 2\delta_c(k) +\delta_i(k))}}{r^2} dr \nonumber \\
\label{XAFS_eq}
\end{eqnarray}

\begin{equation}
A_i=N_i S_{0}^2,
\label{A_i}
\end{equation}

\noindent where $g_i$(r$_{0i}$,r) is the i$^{th}$ shell pair distribution
function (PDF) for atoms at a distance r$_{0i}$ from the center atom, (here
Mn), $F_i(k,r)$ is the back scattering amplitude, and $\delta_c(k)$ and
$\delta_i(k)$ are the phase shifts from the central and backscattering atom
potentials respectively.  The amplitude, A$_i$ (Equ. \ref{A_i}), is the product of the
coordination number, N$_i$, from diffraction and S$_0^2$, the
amplitude reduction factor, which is included to correct for multi-electron
effects since multi-electron processes contribute to the edge step-height but
not to the EXAFS amplitude. Experimentally, S$_0^2$ also corrects for
several other small effects such as small errors in the estimation of the mean
free path in the theoretical calculations, a small amplitude reduction in
the data because of the X-ray energy resolution, some harmonic content in the
synchrotron beam, non-uniformity/pinholes in powder samples, etc.  Finally an
additional parameter, $\Delta$E$_0$, describes the difference in edge energy
between the value defined for the data (half height) and the theoretical
functions (for which $k=0$ at E$_0$).

Since the EXAFS measurements were collected on two different beamlines with
different energy resolutions and with two different set-ups for the low and
high-T ranges, there will be slight variations in the amplitude reduction
factor (and we have to choose two slightly different S$_0^2$ for each set of
data).  Further discussion about setting S$_0^2$ will be given later.
 
In fitting the EXAFS data, we assume a Gaussian PDF with a width $\sigma$ for
the first shell Mn-O peak.\cite{Downward2005} Also note that different
contributions to $\sigma^2$ add up in quadrature if the different distortion
mechanisms are uncorrelated; i.e. $\sigma_{total}^2$ = $\sigma_{phonons}^2$ +
$\sigma_{static}^2$ + $\sigma_{polarons}^2$ + $\sigma_{J-T}^2$.  According to
the model proposed by Downward {\it et al.}\cite{Downward2005}, near T$_c$ a
fraction 2y of the sites are covered by two-site polarons (dimerons)
($\sigma_{polarons}^2$) and the remaining (1-2y) are J-T distorted electron
sites ($\sigma_{J-T}^2$).  These two contributions are summed to give
$\sigma_{J-T/polarons}^2$; this quantity will vary with magnetization (and
temperature) as first the dimerons, and at lower T, the more distorted J-T
sites, become magnetized.

\section{Magnetization and resistivity data} 

\begin{figure}
\vbox{
\includegraphics[width=3.0in,angle=0]{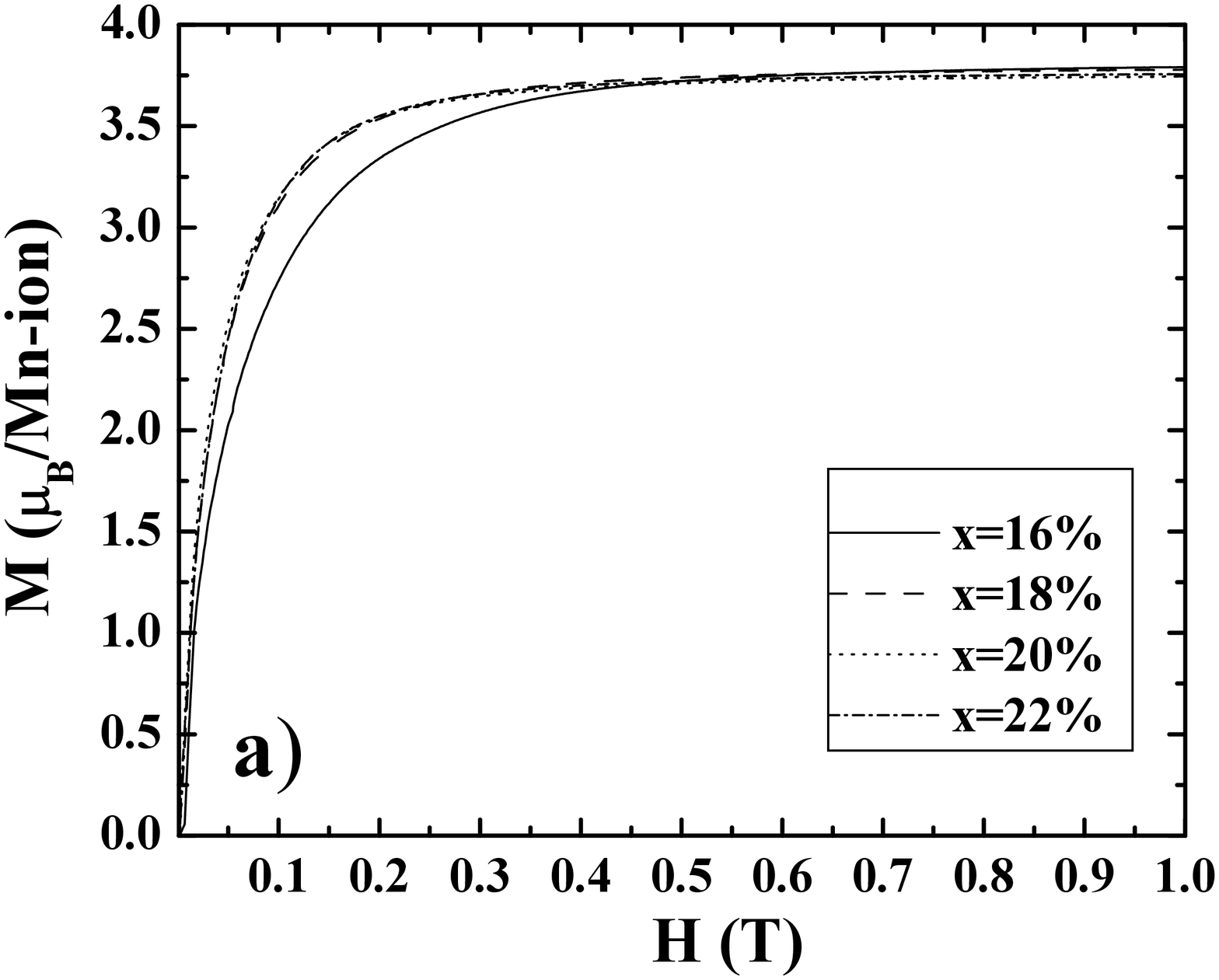}
\includegraphics[width=3.0in,angle=0]{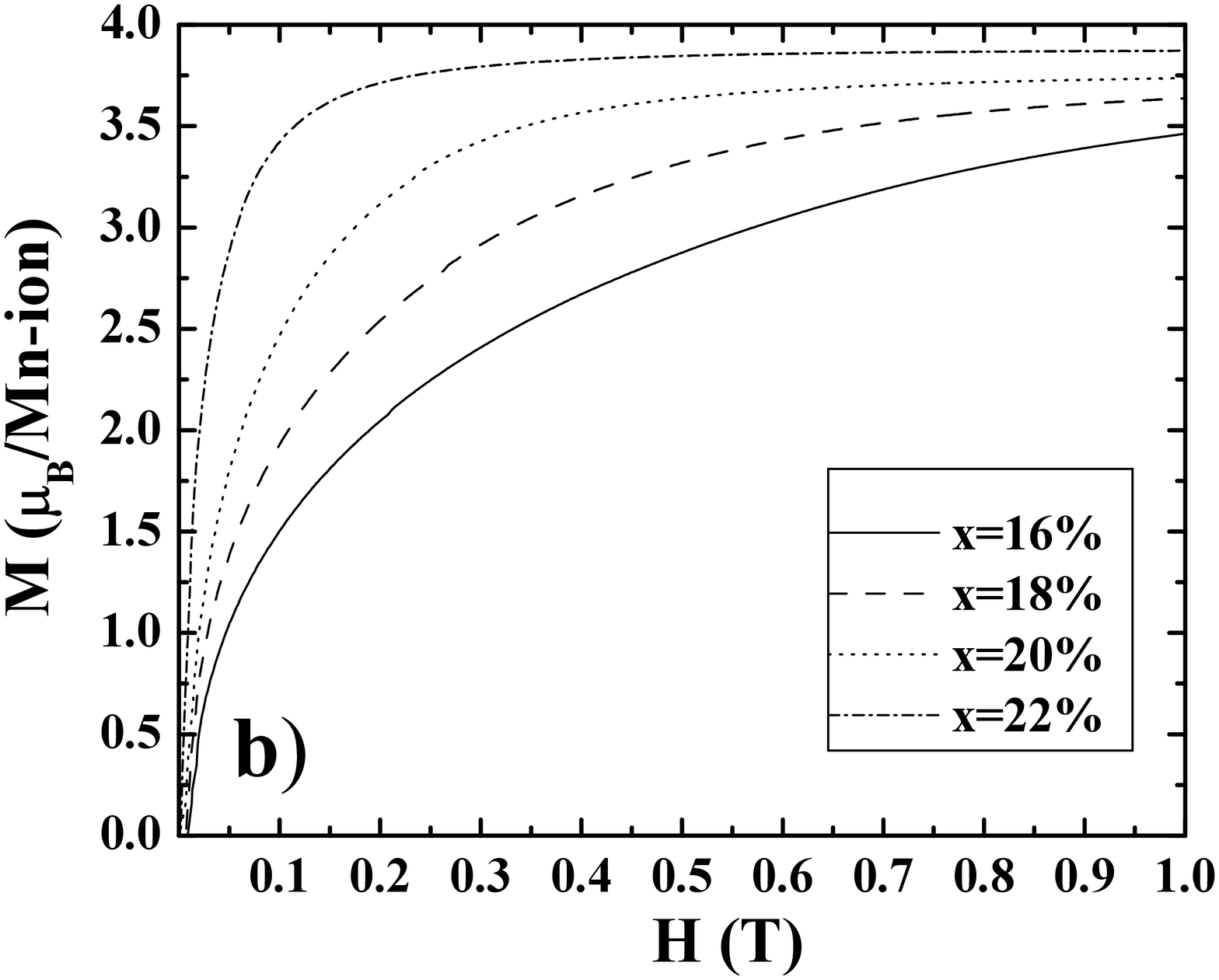}
\includegraphics[width=3.0in,angle=0]{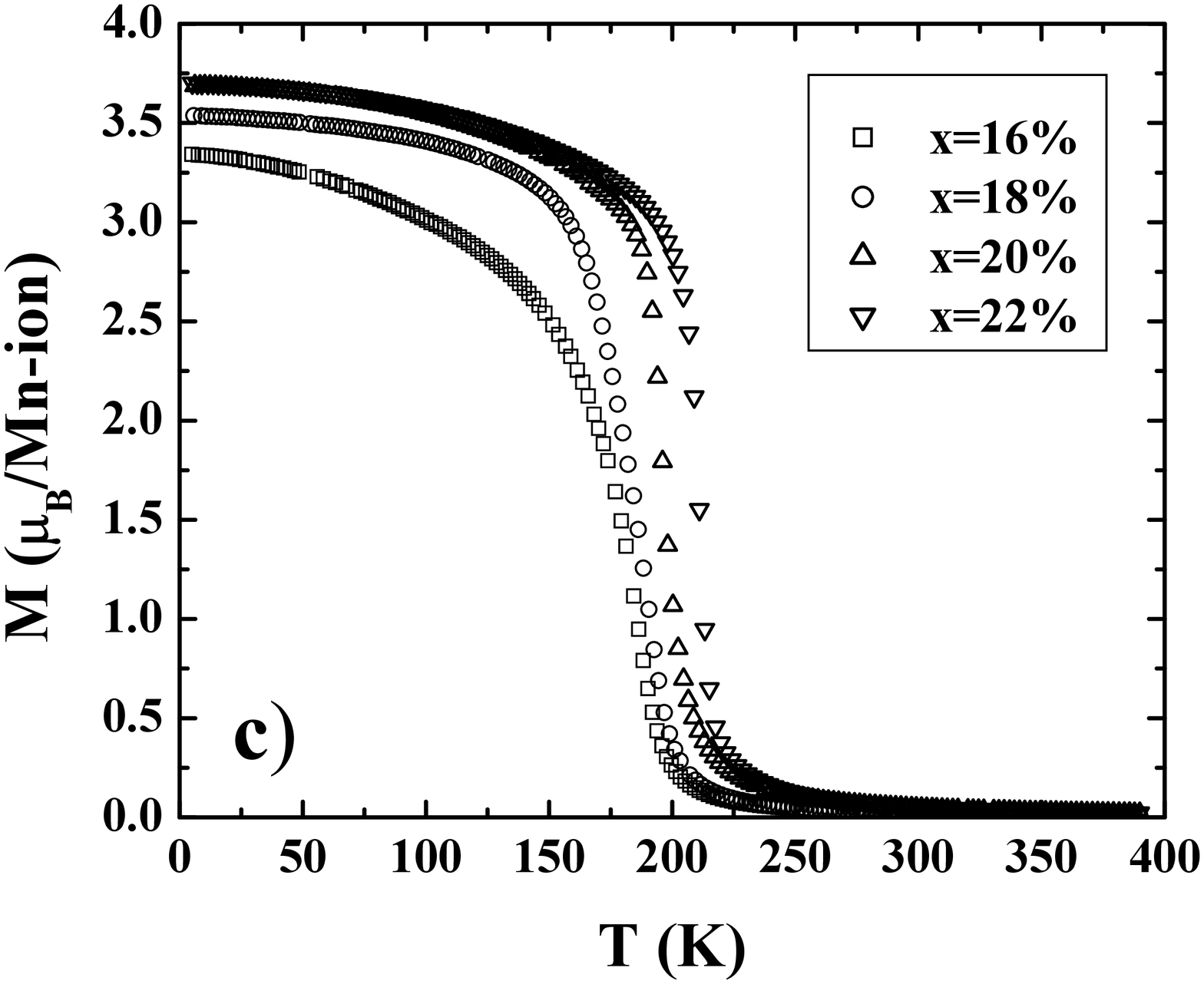}
\includegraphics[width=3.0in,angle=0]{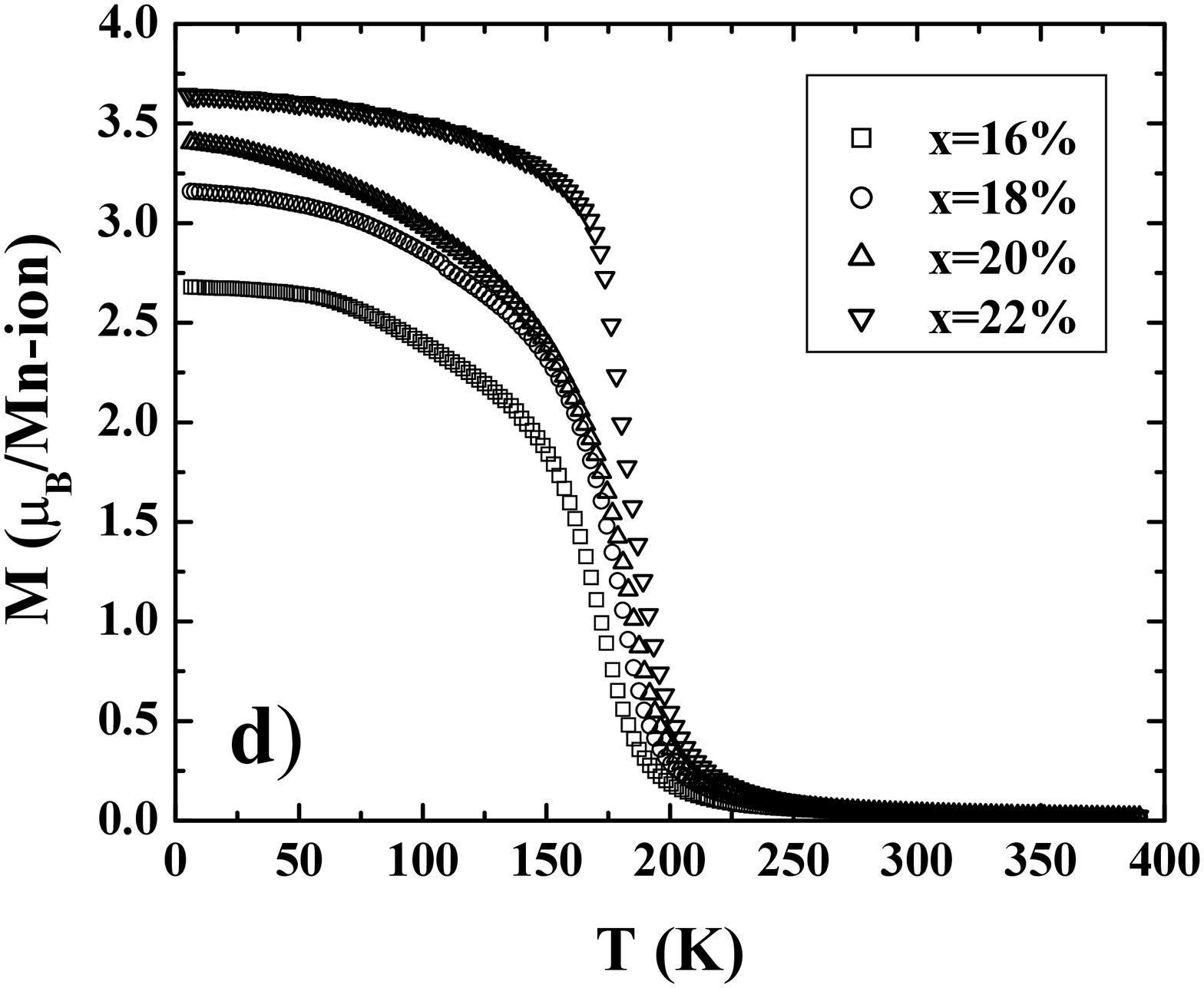} }

\caption{ Magnetization as a function of B-field and temperature:  a) M vs B for the
as made samples (5K), b) M vs B for the annealed samples (5K), c) M vs T for
 the as-made samples (0.4T for x = 0.2, 0.22; 0.2T for x = 0.16, 0.18)
and d) M vs T after annealing in Ar (0.4T).  These figures show that the magnetization is
highest at low T and is fairly close to the theoretically expected value for
the as-made samples.  For the annealed samples the saturation magnetization is
smaller, and significantly reduced for the 16\% Ca sample at 0.4T. Only at high
B-fields does M approach the saturated value. The number of data
points plotted is reduced to 200 for each curve for clarity. }

\label{mvst}
\end{figure}

The magnetization, M, is plotted as a function of B in Fig. \ref{mvst}a,b for
the as made and Ar annealed samples respectively. These plots show the
evolution of the magnetization process as th hole concentration is lowered. The
as made samples have reached saturation at or below 0.4T as observed previously
for CMR samples.\cite{Alonso00,Downward2005} However for the Ar annealed samples
the curves are spread out over a range of B-fields. Although the x = 0.22 sample
also saturates below 0.4T, the other samples significantly larger magnetic fields to reach saturation, and M is still increasing
at 1T for the x = 0.16 and 0.18 samples. An important question here is whether
the lower magnetization at 0.4T means that some sites are not yet magnetized or
that all the Mn spins are magnetized but some sites are in non-aligned clusters.
We argue that it is likely the former. First for all CMR samples the DE coupled
FM sites are easily aligned at 0.4T and the same should apply for the main FM
domains for slightly lower Ca concentrations. However as the Ca concentration
decreases there will be an increasing number of tiny regions with no Ca, and
hence no local holes.  These nanoscale regions, of order a few unit cells, will
possess mostly AFM coupling between a few Mn sites (but no long range AFM order)
since that is the dominant magnetic coupling when no holes are present. In
addition there will be Mn spins on the boundary between these AFM-coupled
nanoclusters and the large FM clusters. Such spins may be frustrated - having
AFM coupling to the nano AFM cluster but FM coupling to the FM clusters as a
result of occasional hole hopping onto these sites.   Consequently the boundary
Mn sites and the AFM coupled nano-clusters may not be aligned at low B fields,
but can be forced to align at high B-fields. The structural results obtained
from the EXAFS results below supports this scenario and will be discussed in more detail 
later.

Fig. \ref{mvst}c,d shows the magnetization, M, as a function of T before ({Fig.
\ref{mvst}c}) and after (Fig. \ref{mvst}d) the anneal in Ar to reduce the O
content - the Ca and final hole concentrations are x = 0.16, 0.18, 0.2, and
0.22, and y = 0.2, 0.22, 0.23, and 0.25 respectively.  Note that the values of
T$_c$ are reduced by 15-20K after the anneal for each concentration.  For each
plot the saturation magnetization decreases as the Ca concentration decreases
whereas a very slight increase is expected ({M$_{sat}$ = (4-x)$\mu_B$}); more
importantly, the saturation magnetization in the annealed samples is
significantly reduced at 0.4T as discussed above. For the as-made samples the
saturation magnetization at low T in these plots is low for the x = 0.16 and
0.18 samples because these data were collected at 0.2T instead of 0.4T prior to
annealing.  For the Ar annealed samples, the saturation magnetization for the
16\% Ca sample drops by $\sim$30\% at 4K (B = 0.4T) compared to the 22\% sample
(Fig. \ref{mvst}d).  

\begin{figure}
\vbox{ 

\includegraphics[width=6.0in,angle=0]{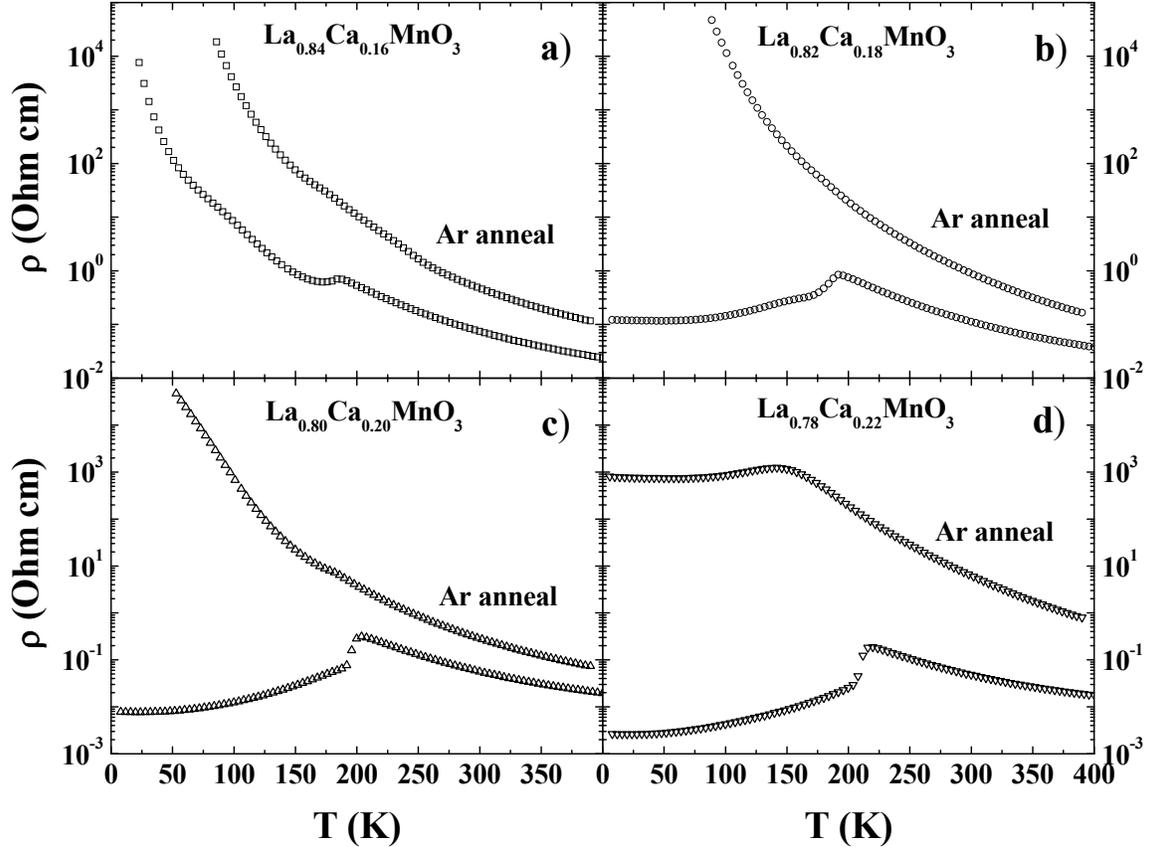}
}

\caption{Plots of the resistivity before and after Ar annealing for the four
samples. For the as-made samples, the M/I occurs between x = 0.16 and 0.18; for
the Ar annealed samples, only the x = 0.22 sample shows metallic-like behavior
at low T. The number of data points plotted has been reduced to 100 for each
curve. }

\label{rhovst}
\end{figure}

In Fig. \ref{rhovst}, the resistivity is plotted for the as-made (Fig.
\ref{rhovst}a) and annealed (Fig. \ref{rhovst}b) samples.  For the as-made
samples, the M/I transition occurs between 0.16 and 0.18 Ca concentration, and
$\rho$(T) varies significantly near T$_c$. After the reduction of oxygen via
the Ar anneal, the resistivity increases considerably and only the 22\% Ca
sample shows a M/I transition.  From the above data, the annealed samples are
mostly in the FM insulator regime and thus are the more important samples for
this study. Note that the small reduction in O content has a significant change
in the bulk magnetization for the x = 0.16 sample.


\section{EXAFS Data and Analysis}
\label{data}

The EXAFS data were reduced using the RSXAP package,\cite{RSXAP} which
implements standard data reduction techniques.  A pre-edge background was
removed from the data (the Victoreen formula was used to adjust the slope above
the edge after the pre-edge subtraction)  and an experimental E$_0$ was defined
as the energy of the half-height point on the Mn K-edge.  The post-edge
background was removed using a spline with five knots to approximate $\mu_0$ in
$\mu(E)=\mu_0(1+\chi(E))$.  The background-subtracted data  $\chi$(E) were then
transformed to k-space using the relation $k=\sqrt{\frac{2m(E-E_0)}{\hbar^2}}$.  

\begin{figure} 
\includegraphics[width=4.0in,angle=0]{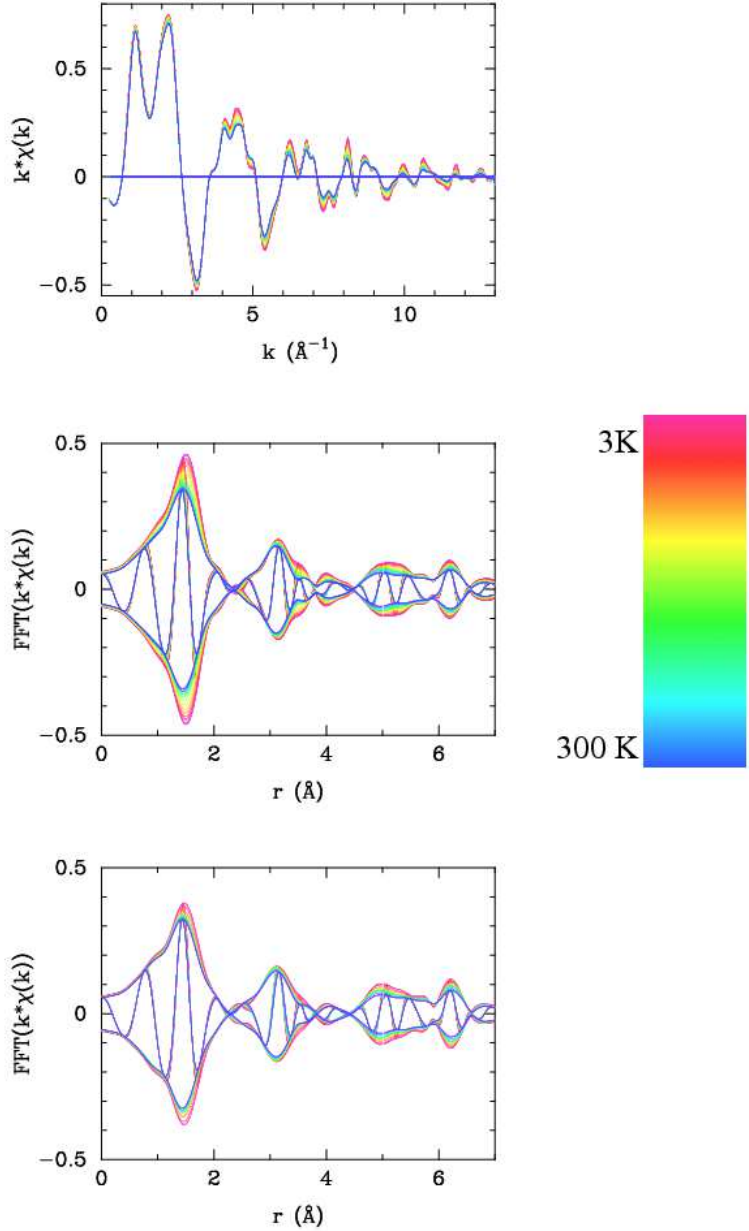}  

\caption{(Color online) A comparison of EXAFS data for the x = 0.16 and 0.22
samples after annealing (hole concentrations are y = 0.20 and 0.25
respectively). Top:  k-space data for x = 0.22, Middle: r-space data for x =
0.22; Bottom: r-space data for x = 0.16.
In each case, the amplitude is highest at low temperature and decreases
monotonically with increasing temperature. Note that the amplitude for the x =
0.16 sample (Bottom) is smaller overall (more disorder) and has a weaker T
dependence compared to the x = 0.22 sample (Middle).  } 

\label{ks_rs} 
\end{figure}

Next the k-space data k$\chi(k)$ were Fast Fourier Transformed (FFT) to r-space
with a k-space window of 3.3-12.0 \AA$^{-1}$ with a Gaussian broadening of
width 0.2 \AA$^{-1}$.  Examples of the data are shown in Fig. \ref{ks_rs} for
two annealed samples (x = 0.16 and 0.22). 

The data were then fit to theoretical EXAFS functions generated by FEFF 8.20
(developed by Rehr and co-workers\cite{FEFF8}), using the program \verb|rsfit|
(RSXAP package).  Our primary interest here is the width of the Mn-O PDF which
parametrizes the amount of distortion present.  We used only one value of
$\sigma$ for the Mn-O shell as in previous work,\cite{Downward2005} and fixed
the number of oxygen neighbors to the lattice structure (N$_1$ = 6 neighbors)
for each sample.  To get a reasonable fit, a number of other constraints on the
parameters are also required.  First, for data collected on the same sample and
on the same beamline, $\Delta$E$_0$ was obtained by allowing $\sigma$ and
$\Delta$E$_0$ to vary on the lowest temperature data; the average value
obtained from those fits was used to constrain $\Delta$E$_0$ for the rest of
the data in the same set.  For the parameter S$_0^2$, a number of fits were
carried out.  In the first set of fits, we let the amplitude A$_1$ (A$_1$ =
N$_1$ S$_0^2$) vary for the low temperature data and determined S$_0^2$ from
those fits.  Then A$_1$ was kept constant for fits as a function of T.  The
main effect of small changes in this parameter is a vertical shift of plots of
$\sigma^2$ versus T.  Once S$_0^2$ for the low temperature data (3-300 K) was
determined, we slightly adjusted S$_0^2$ for the high temperature data set
(300-550K) to make the values of $\sigma^2$ align with the low T data at 300K.
For all the data below 330K, S$_0^2$ $\approx$ 0.75; the variation to join high
T and low T data is $\approx$ 0.01.  

\begin{figure}
\vbox{
\includegraphics[width=3.2in]{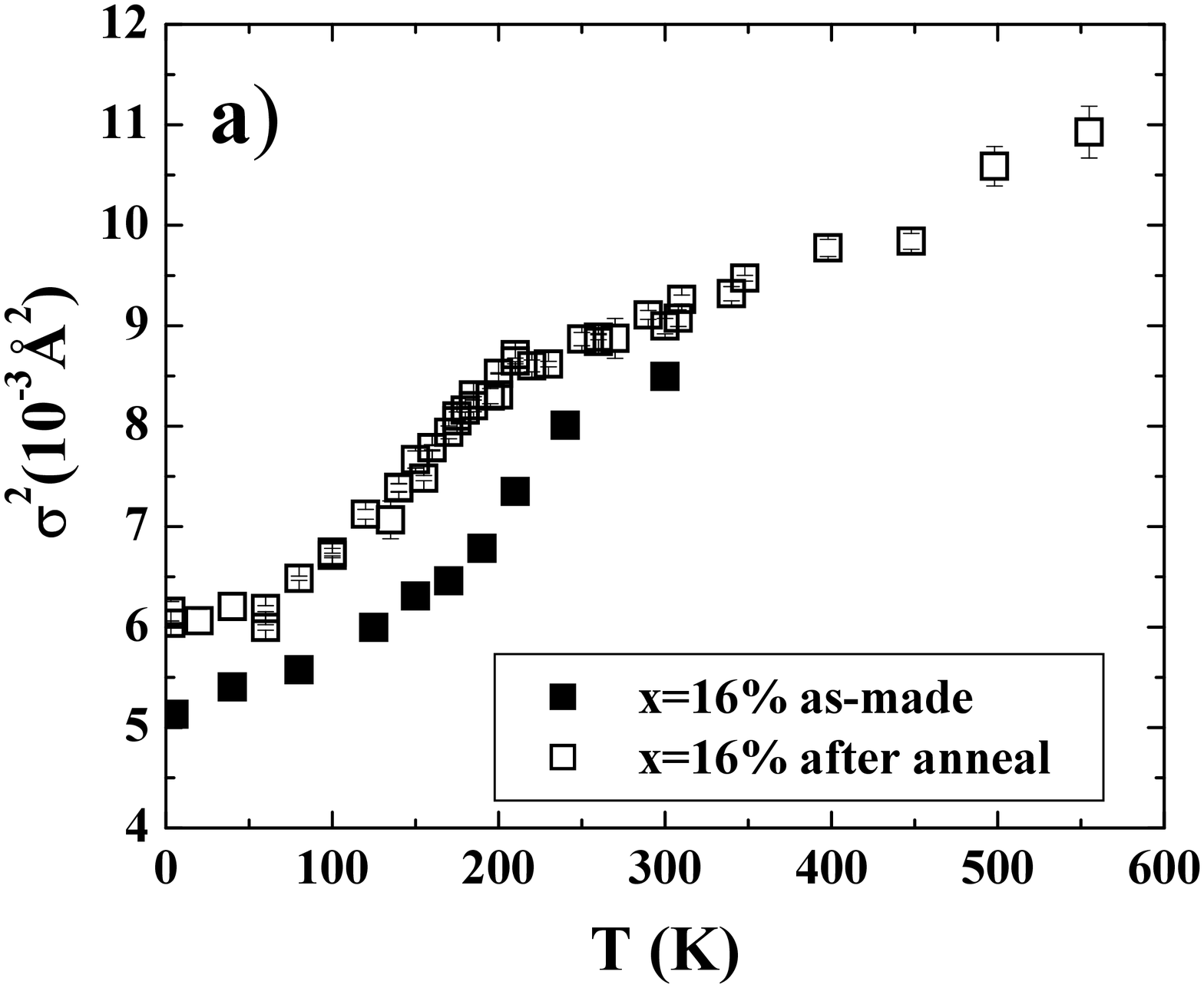} 
\includegraphics[width=3.2in]{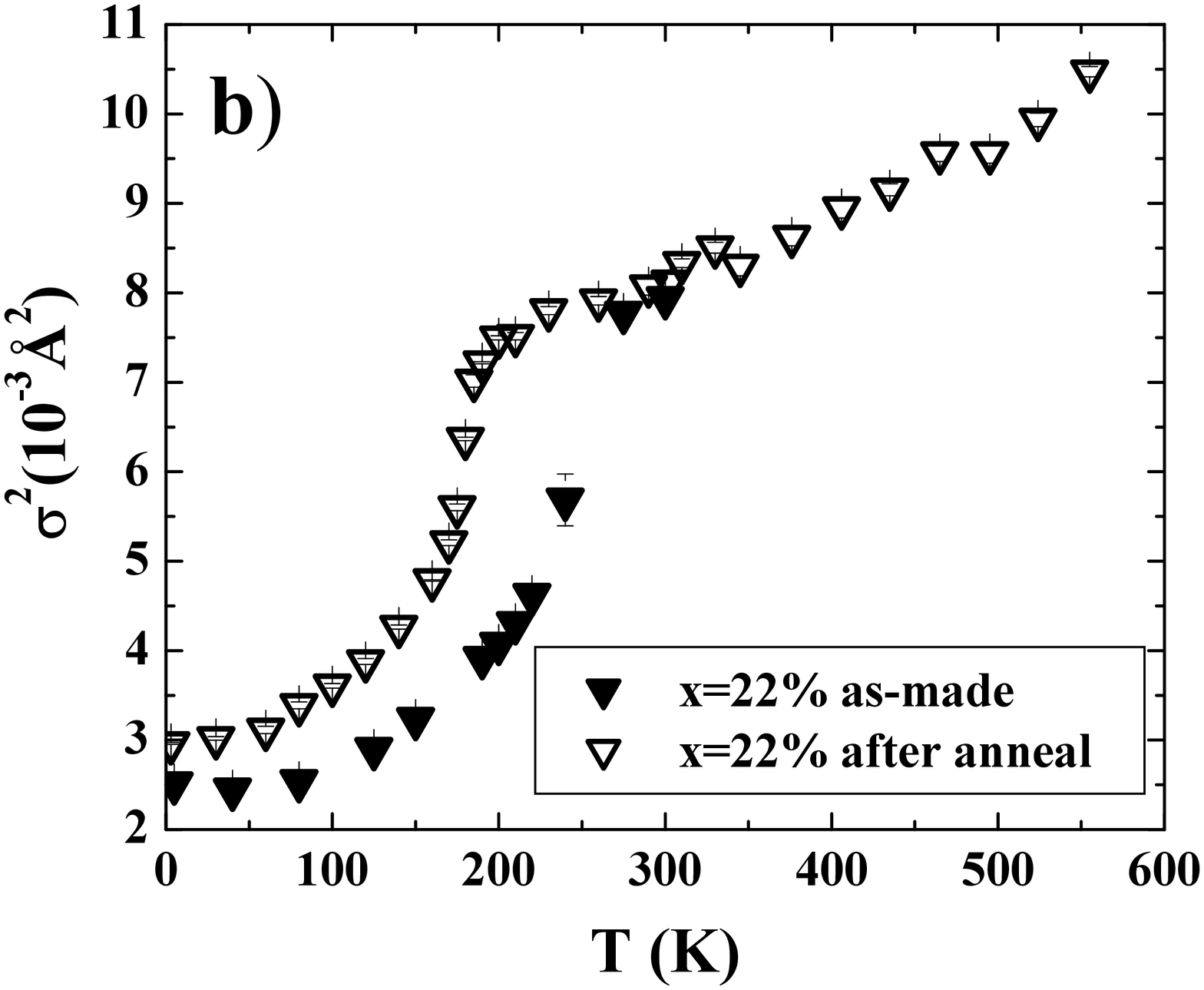}
}

\caption{ $\sigma^2$ as a function of T for: (a) x = 0.16; and (b) x = 0.22
samples, before and after the Ar anneal. The anneal shifts the step in
$\sigma^2$ (near T$_c$) to lower temperatures and the height of the step is
reduced. Note the different vertical scales used for (a) and (b). }

\label{com_anneal}
\end{figure}

Most of the data were collected on the Ar annealed samples but a few points
were collected on two of the as-made samples. In Fig. \ref{com_anneal},
$\sigma^2$ is plotted (on different expanded scales) as a function of T before
and after the anneal for x = 0.16 and 0.22. Although the data for the as-made
samples are sparse (solid points), it is clear that after annealing there is a
shift of T$_c$ to lower T for both samples (See also Figs. \ref{mvst} and
\ref{rhovst}) and that the step-height decreases, particularly for the x = 0.22
sample. 

\begin{figure}
\includegraphics[width=3.2in]{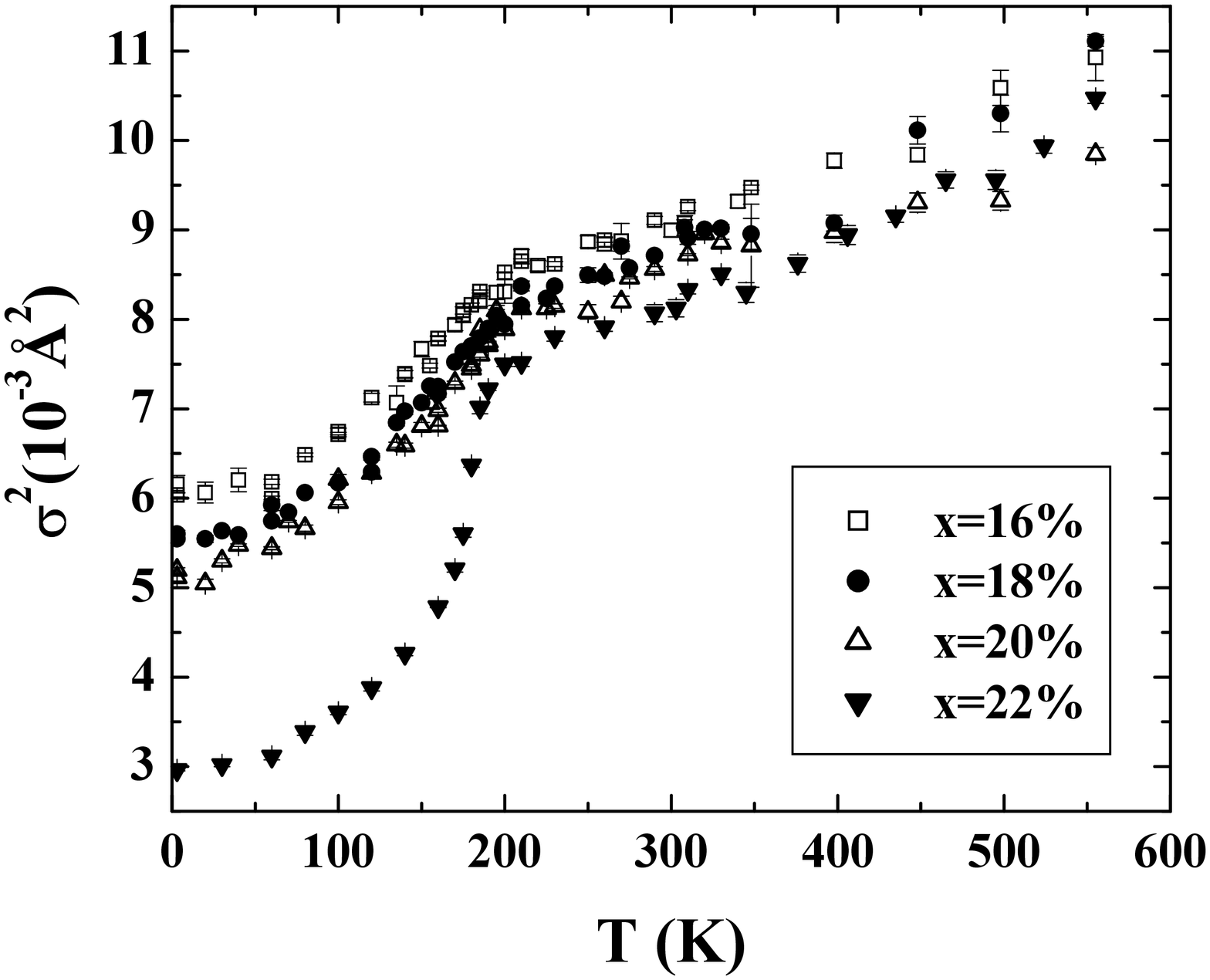} 

\caption{ $\sigma^2$ as a function of T for the four annealed samples (x = 0.16
- 0.22) on the same vertical scale. The plot of the x = 0.22 sample is similar
  to previous samples which show a M/I transition;\cite{Downward2005} however,
the other three samples, which all show insulating behavior and a reduced
saturation magnetization at low T, have a significantly smaller step in
$\sigma^2$ and a slightly increased distortion above T$_c$ (higher $\sigma^2$)
than other CMR samples. 
}

\label{sig2}
\end{figure}


Fig. \ref{sig2} shows $\sigma^2(T)$ for all the annealed samples on the same
scale.  The step increase in $\sigma^2$ is sharp near T$_c$ for the x = 0.22 (y
= 0.25) sample and has a comparable (slightly smaller) step height to that
observed in other CMR samples.\cite{Downward2005} As before, we attribute this
step to a large increase in the distortions of the Mn-O bonds, associated with
dimerons (extended polarons with low distortions) and some more isolated,
highly J-T distorted electron sites, as the temperature is increased through
T$_c$.  However, the other three samples (x= 0.16-0.20) have a much smaller step
near T$_c$, which decrease slightly with smaller x, and a much larger
value of $\sigma^2$ remains at low T. Thus even though a large fraction of the
sample is magnetized (70\% or more), only a small distortion is removed in the
FM phase for these samples.  A significant fraction of the sample (up to $\sim$
30\% for x =0.16) remains distorted at the lowest temperatures (Note:
$\sigma^2$(4K) is well above the value for zero-point motion).  There is also
an increased static distortion above T$_c$ for the insulating samples.  

In addition, the temperature dependence of $\sigma^2$(T) above T$_c$ is
comparable for all samples. The thermal phonon contributions were determined
from a fit of $\sigma^2(T)$ above T$_c$ (205-550K) to the correlated Debye
model (Equ. \ref{cD}) plus a static off-set.  This model is usually a good
approximation for all phonon modes\cite{Ashcroft76} including acoustic and
optical phonons and is given by: \cite{Lee77,Teo86,Bianconi88}

\begin{equation}
\sigma^2_{cDebye} = \frac{3\hbar}{2M_R} \int_{0}^{\omega_D}\frac{\omega}{\omega^3_D}C_{ij} 
coth(\frac{\hbar\omega}{2k_BT}) d\omega;
\label{cD}
\end{equation}

\noindent where $\omega_D$ is the Debye frequency, C$_{ij}$ is a correlation
function given by 1-sin($\omega r_{ij}$/c)/($\omega r_{ij}$/c),
$c=\frac{\omega_D}{k_D}$ where $k_D$ is the Debye wavelength, and
$\sigma^2_{cDebye}$(T$\sim$0) with zero static offset gives the zero-point
motion value of $\sigma^2$.  The slope of $\sigma^2$$_{cDebye}$(T) vs T is very
low at low T and increases to a constant value (determined by the spring
constant, reduced mass, and C$_{ij}$) for T $>$ $\Theta_D$. See Ref.
\onlinecite{Teo86} for details. 

\begin{figure}
\vbox{ \includegraphics[width=3.2in]{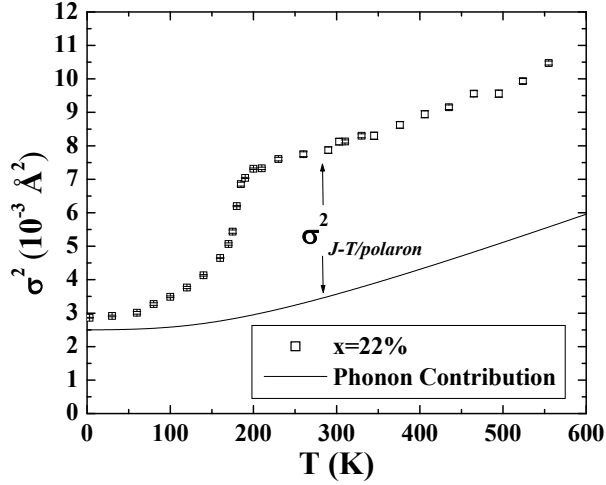}
}

\caption{$\sigma^2$ vs T for LCMO, x = 0.22. The solid line is the phonon
contribution calculated using the correlated Debye model with zero static
distortion and $\theta_D$ = 830 K ($\theta_D$ for all the samples above
T$_c$ varies from 812 K to 860 K with an average of about 830 K). 
$\sigma^2_{J-T/polaron}$(T) - the J-T/polaron contribution - is defined as the
difference between the experimental data and the phonon contribution at each
temperature point.}

\label{debyefig} 
\end{figure}

We obtained values of $\Theta_D$ $\sim$ 812-860K $\pm$ 30K for the four samples
with an average of about 830K. This agrees well with the value $\Theta_D=860$ K
for the Mn-O bond in Ca-substituted LaMnO$_3$ materials as obtained
previously.\cite{Booth98b,Mannella2004,Downward2005} This fit passes through
the points above T$_c$ (T$>$200K) very well. In Fig. \ref{debyefig}, we plot
the results for x = 0.22 and the Debye curve ($\Theta_D$ = 830K) with the
static contribution removed. This represents the thermal contribution to
$\sigma^2$ (including zero-point-motion); notice that this curve is parallel to
the data above T$_c$.


\begin{figure}
\vbox{
\includegraphics[width=6.5in]{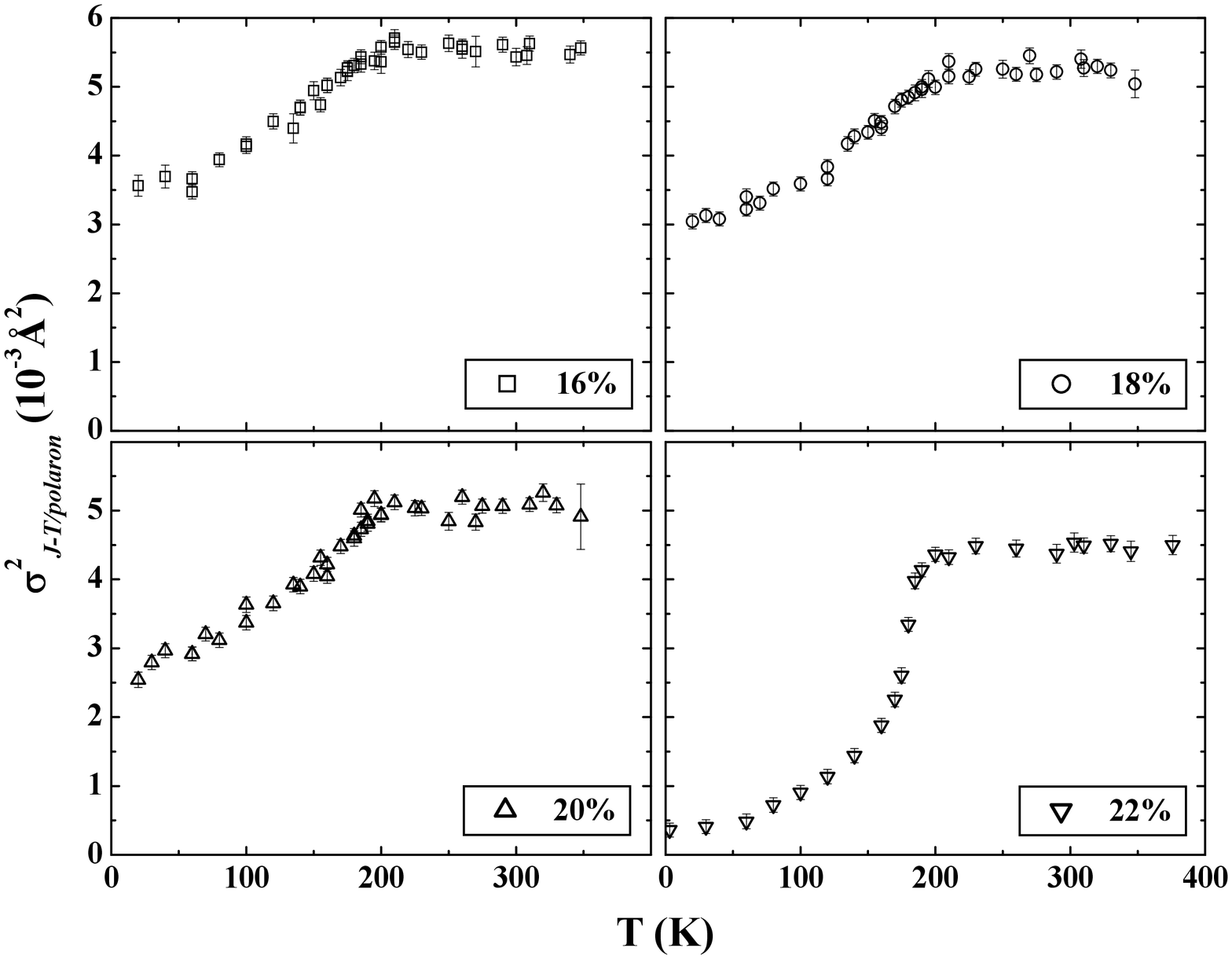}
}

\caption{$\sigma^2_{J-T/polaron}$(T) vs temperature for LCMO x=16-22\%.  The
non-thermal contribution, $\sigma^2_{J-T/polaron}$(T), is defined in Fig
\ref{debyefig}.  Note that above T$_c$ the $\sigma^2_{J-T/polaron}$ is
independent of T, and that as the concentration decreases the remaining
distortion at low T increases.}

\label{lcmodsig2vst}

\end{figure}

The non-thermal-phonon contribution to $\sigma^2$ that is removed as the sample is
cooled below T$_c$, $\sigma^2_ {J-T/polaron}$(T), can be obtained by subtracting
the Debye curve (solid line in Fig. \ref{debyefig}) from the $\sigma^2$(T) data
for each sample. The results are shown in Fig. \ref{lcmodsig2vst}. This plot
illustrates several important aspects of $\sigma^2_{J-T/polaron}$(T).  First,
above T$_c$ it is independent of T - the polaron and J-T distortions are fully
formed and do not change significantly with T. Second, the step decrease as T
goes to zero is greatly diminished for the insulating samples, although the
samples are still highly magnetic (see Fig. \ref{mvst}).  Third (and related to
the second point), the remaining static distortion increases as the Ca
concentration is lowered, and as the fraction of sample that remains
unmagnetized at 0.4T, increases.

\begin{figure}
\vbox{
\includegraphics[width=6.5in]{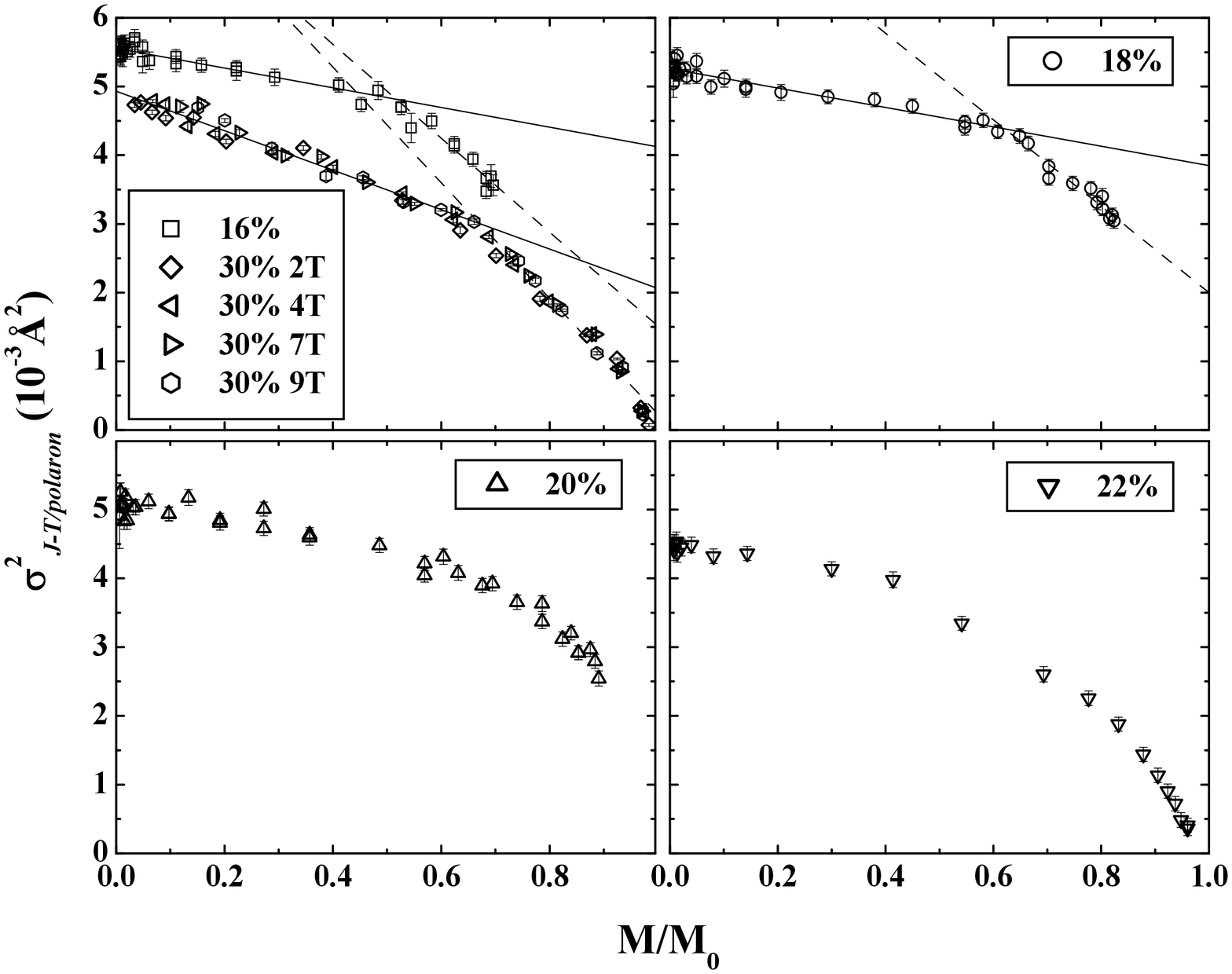}
}

\caption{$\sigma^2_{J-T/polaron}$ vs $\frac{M}{M_0}$ for LCMO x=16-22\%.  Note
that the slope at low M is comparable for all samples and that the J-T/polaron
distortion is only partially removed for the x = 16-20\% samples. For
comparison, the data for the 30\% sample from Downward {\it et
al.}\cite{Downward2005} are replotted in this format below the results for
the 16\% sample. }

\label{lcmodsig2vsmag}

\end{figure}


To investigate the correlations between local structure and magnetization, we
plot the data in a new way\cite{Downward07} which simplifies the discussion.
By combining $\sigma^2_{J-T/polaron}$(T) (Fig. \ref{lcmodsig2vst}) with the
M(T) data (Fig. \ref{mvst}) we plot $\sigma^2_{J-T/polaron}$(T) vs
$\frac{M}{M_0}$ for the four samples in Fig. \ref{lcmodsig2vsmag}. Here we use
the fraction $\frac{M}{M_0}$ ($M_0$ is the theoretical saturation
magnetization) as an approximate measure of the fraction of magnetized states.
At low M, $\sigma^2_{J-T/polaron}$ decreases slowly with M, i.e. the distortion
removed per Mn site is very small until the sample is at least 50\% magnetized.
At higher magnetization, $\sigma^2_{J-T/polaron}$ drops much more rapidly with
M and approaches zero for the 22\% sample. However for the insulating samples,
relatively little distortion is removed overall. Note that the slope at low M
is similar for all samples, but the 20\% and 22\% samples have a continuous
change in slope. 
Here we need to point out one caveat - the EXAFS data were collected at B = 0T,
while the magnetization data were collected at B = 0.4T. However, we note that
while this field is large enough to be reflective of the bulk, zero-field,
magnetization, it is small enough to have a minimal effect on the
magnetoresistance.  

Recently Downward {\it et al.}\cite{Downward2005,Downward07} have shown using
EXAFS measurements as a function of magnetic field for four samples from
21-45\% Ca, that a universal relationship between the local distortions and the
magnetization only exists for high applied B-fields e.g. B $>$ 2T. At much
lower B fields, domain effects are important and the observed bulk
magnetization is lower than expected for the number of magnetized sites implied
from the EXAFS data. 
This analysis suggest that the steeper the M(T) plot, the more curved the
$\sigma^2_{J-T/polaron}$(T) vs $\frac{M}{M_0}$ plot becomes. For the samples
considered here, M(T) for the 16 and 18\% samples have a slower T dependence
than the other samples. It is only for these samples that a clear break point
near $\frac{M}{M_0}$ $\sim$ 0.5-0.6 is present in Fig. \ref{lcmodsig2vsmag}, as
observed previously for the higher Ca concentrations.\cite{Downward2005}

Surprisingly the magnitude of the slope at low M for the 16\% Ca sample (0.0014
\AA$^2$) is a factor of two lower than for the 30\% Ca sample.  For comparison,
the latter data\cite{Downward2005} are replotted here, below the data for the
16\% sample ( Fig. \ref{lcmodsig2vsmag}); this indicates that the initial
magnetized sites have even less distortion than for the CMR samples. 
Note that the high M slopes
for the low Ca concentration samples (x = 0.16, 0.18) are comparable (within
25\%) to that for the CMR samples. See Table \ref{table} for a comparison of
the high and low slopes with earlier results for CMR samples x = 0.21 and 0.3.

One possibility for the smaller low slope for the FMI samples is that the holes
are confined to a smaller number of sites and hop rapidly between only a few Mn
atoms, keeping them relatively undistorted. A major difference for this 16\% Ca
sample is that the magnetization only reaches M/M$_0$ $\sim$ 0.7 at low T and B
= 0.4T. If one extrapolates the straight line through the data from
$\frac{M}{M_0}$ = 0.5 to $\frac{M}{M_0}$ = 1.0, the distortion that would have
been removed if every site became magnetized via DE, is nearly a factor of 2
larger (see dotted line for the 16\% Ca sample in Fig. \ref{lcmodsig2vsmag}).
The remaining J-T/polaron distortion at $\frac{M}{M_0}$ = 1.0, $
\sigma^2_{J-T/polaron}$ $\sim$ 1.5x10$^{-3}$ \AA$^2$, is comparable to the
excess distortion observed above T$_c$ (Fig. \ref{sig2}) within our
uncertainties (in comparison, $\sigma^2_{J-T/polaron}$ $\sim$ 0 for the
30\% sample at $\frac{M}{M_0}$ = 1.0).  Thus it is as if the magnetization
process were truncated for the lower concentration samples before all sites
became magnetized.

\begin{table}
\caption{Table of slopes from plots of $\sigma^2_{J-T/polaron}$
vs $\frac{M}{M_0}$.}

\begin{tabular}{|c|c|c|c|c|}
\hline
Concentration  & Low slope (10$^{-3}$\AA$^2$) & High slope (10$^{-3}$\AA$^2$) & breakpoint & chisq/ndf  \\
\hline
16\%  & -1.44 & -6.84 & 0.517 & 0.790 \\
18\%  & -1.41 & -6.29 & 0.621 & 1.061 \\
21\%  & -3.18 & -6.31 & 0.611 & 4.497 \\
30\%  & -2.95 & -8.65 & 0.678 & 2.123 \\


\hline

\end{tabular}
\label{table}
\end{table}


The small overall distortion removed in the FMI phase for x = 0.16-0.2,
provides direct evidence that low distortion (non-magnetized) Mn sites
associated with the charge carriers, exist at T$_c$ and are magnetized first.
These are the dimeron quasiparticles proposed by Downward {\it et
al.}\cite{Downward2005}  This further supports the proposal that there are at
least two types of distorted sites in this system: one associated with the
delocalized hole charge carriers (dimeron or two-site polaron) and another with
the remaining J-T distorted Mn sites.


\section{Discussion and conclusions}
\label{discuss}

The first issue to discuss is the structure of these systems above T$_c$.
Neutron diffraction studies show that between 200 and 300K the samples become
pseudo-cubic and the Mn-O bond lengths are nearly equal above a temperature
T$_{J-T}$, at these Ca concentrations.\cite{Biotteau01,Pissas05}  In contrast
$\sigma^2_ {J-T/polaron}$ from  EXAFS shown in Fig. \ref{lcmodsig2vst} has a
large distortion of the Mn-O PDF above T$_c$, and it remains constant up to
400K.  Biotteau {\it et al.}\cite{Biotteau01} ascribe this transition to a
change from a quasi-static to a dynamic J-T effect.  We agree with this
assignment - EXAFS is a very fast probe ($\sim$ 10$^{-15}$ s) and can follow
any structural J-T distortions.  The constant value of $\sigma^2_
{J-T/polaron}$(T) above T$_c$ means that the magnitude of the J-T distortion is
unchanged. The collapse of the Mn-O PDF observed in neutron diffraction means
that the lattice fluctuations associated with J-T distortions become faster
than the time scale of the neutrons ($\sim$ 10$^{-12}$ s) above T$_{J-T}$. This
underscores the importance of including dynamics in any discussion of these
systems.

The main results reported above for samples in the FMI regime are: 1)
the functional relationship between changes in the non-thermal contributions to
$\sigma^2$ ($\sigma^2_{J-T/polaron}$(T)) and the sample magnetization are the
same as in the FMM regime; 2) The overall distortion that is removed as
the sample becomes ferromagnetic at low T is significantly  smaller for the
FMI compared to the FMM, even for a rather small change in Ca
concentration; and 3) A significant fraction of the sample (5-30\% in this
concentration range) remains unmagnetized and is still distorted at low
T (4K) at B = 0.4T.

Point \#1 suggests that the basic mechanism for FM behavior which involves the
spins, the mobile charges, and the lattice, is the same in both the insulating and
metallic FM regimes- thus the DE model (plus local distortions) is also the dominant 
coupling mechanism for most of the Mn sites at low Ca concentrations. Point \#
2 may at first be surprising but is consistent with the earlier extensive
results reported by Downward {\it et al.}\cite{Downward2005} for samples in the
CMR regime, namely that there is a large fraction of sites associated with the
charge carriers that have a small distortion per Mn site. The small overall
distortion removed for the x=0.16-0.2 samples is further evidence of this
small distortion/site.\cite{Downward2005} 

Point \#3 may also be surprising - how does a small fraction of the sample
consistently form an insulating layer between more conductive regions? It is
perhaps less surprising if one first looks back at the LCMO CMR systems at
partial magnetization. From our previous work, the turnover of the resistivity
occurs when a significant magnetization has developed;
up to that value of magnetization the resistivity is still increasing as T
decreases (i.e. an insulator or semiconductor); also the turnover value of
magnetization increases as the Ca concentration decreases.  This behavior is
comparable to the FM insulator sample with x = 0.16, except that turnover does
not occur.  The difference is that for the CMR sample the most distorted
fraction of the sample does become magnetized at lower T - and the sample
becomes conducting.  For the FMI, much of this fraction remains distorted and
never becomes magnetized at 0.4T.  Thus the insulating and metallic systems are
quite similar at partial magnetization.  

It is useful here to comment further on the issue as to whether at 0.4T the
sample is only partially magnetized or is fully magnetized but many domains are
not aligned.  Fig. \ref{lcmodsig2vsmag} shows that at low M the first sites to
be magnetized have a low distortion per site. We argue that these sites are
related to the DE coupled FM clusters. If one argues that all sites are
magnetized at 0.4T, then the data for the 16\% sample in Fig.
\ref{lcmodsig2vsmag} should  range up to  $\frac{M}{M_0}$ = 1.0 instead of 0.7, 
the break point would be near 0.75; then there
would be an even larger fraction of easily magnetized sites with low
distortions which would be inconsistent with the large number of filled e$_g$
sites at low Ca concentrations. The only alternative would be to invoke a FM
interaction that would lead to easily magnetized sites at low M that remain
highly distorted (and thus have little distortion removed in the magnetized
state). This seems unreasonable in view of the properties at slightly higher
Ca concentrations. 

In their dimeron model, Downward {\it et al.}\cite{Downward2005} suggested that
the dimerons are preferentially located along filamentary clusters close to Ca
sites; the argument is based on charge neutrality - the dimerons are + charged
holes delocalized over two Mn sites while Ca on a La site acts like a negative
charge.  In such a model the chains of linked unit cells containing a Ca atom
form pseudo-one-dimensional filaments within the insulating LaMnO$_3$ host.
Once the dimeron sites are magnetized as filamentary clusters, a further
increase of magnetization requires magnetizing the surrounding J-T distorted Mn
sites of the host lattice. Once enough of these sites are magnetized it will
connect the filamentary chains and make the sample conducting; however the
smaller the fraction of the sample that forms the filamentary structures, the
larger the number of distorted Mn$^{+3}$ (e$_g$) sites that would need to be
magnetized (and be kept undistorted via charge hopping) before the filaments
are connected and metallic conductivity can be achieved.

Note that these results again point to an intrinsic inhomogeneity of the
material at the level of a few unit cells. One needs to consider nanoscale
regions with quite different distortions, that change with the local
magnetization - which in turn is determined by changes in T or B. These regions
are determined by the distribution of dopants such as Ca, and also on the
location of metal atom (Mn, La etc.) vacancies. It is likely that the poorly
defined concentration at which the M/I occurs in LCMO  is due to several
connected effects -- the distribution of Ca, correlations between the location
of Ca sites and the location of vacancies and the dynamics of the charge
carriers.  The NMR results mentioned in the introduction\cite{Papavassiliou00}
suggest that the low-distortion magnetic and high-distortion non-magnetic
regions may be further subdivided into sites with different local B fields but
the relative fraction of such sites is not clear.

Metallic conductivity through the CMR manganites in the FM phase is often
considered to be a percolative problem but it is not simple percolation.  If
the holes had equal probabilities to be on {\it any} (Mn) site, then there
would be uniform conductivity as is the case for low doping concentrations in
n- and p-type semiconductors - and if such a manganite system were 50\%
magnetized, percolating magnetic clusters would exist across the sample and it
would be metallic. Thus there must be preferred hole sites as discussed above.
Considering only the Ca dopants, at 20 $\pm$2 \% Ca (the range for the
metal/insulator transition) the system is well below the percolation limit for
a cubic crystal (x = 0.31), and even at 25\% - well into in the FM-metallic
regime - the Ca atoms do not percolate; at these concentrations, chains of Ca
atoms form many partially connected filaments but do not have connectivity
across the sample.  The electrical conductivity however depends on the available
sites for the hole quasiparticles which will include Mn sites close to the Ca
dopants. This expands the volume of the sample available for conduction to at
least one unit cell about each Ca atom but still does not guarantee
connectivity as the potentially conductive sites will be fluctuating.  However
if the fraction of conducting sites can be expanded beyond a unit cell (here by
making more of the sample ferromagnetic via hole hopping), then these expanded
filaments will eventually touch and connectivity can be achieved. For samples
close to the metal/insulator transition (either for samples with x $\leq$ 0.2
at low T or for samples in the CMR regime which are only partially magnetized),
the complex conductivity will depend sensitively on the conducting
microstructure of the sample. 

Finally we return to the nature of the Mn sites and the hole quasiparticles.
In the introduction we pointed out that if the holes are hopping rapidly on a
local scale there will be no Mn$^{+4}$ sites on the time scales of most
experiments, but there will be some Mn$^{+3}$ sites if the holes have preferred
locations and some Mn$^{+3}$ sites are rarely visited by a hole. A large
fraction of sites will be an average of occupancy by a hole and by an e$_g$
electron. A hole on a Mn site tends to reduce the local distortion as the J-T
interaction is not present. When a hole moves off a site (and it now contains
an e$_g$ electron) the site will begin to distort since the J-T interaction
becomes active. Thus such sites will have some average distortion depending on
the hopping rate and the time scale of the measurement. EXAFS is a very fast
probe (10$^{-15}$ s), much faster than phonon time scales, and will see a
weighted average of all the distortions in the sample.  Surprisingly the change
in the local distortion is small for the first 50\% of the sample that becomes
magnetized for the CMR samples. This trend continues to be present in samples
through the concentration driven transition from FMM to FMI (x =0.16-0.22). A
crucial point is that the number of low distortion sites is far larger than the
number of holes and involves approximately an equal numbers of electron sites.
This approximate factor of two (1.5-2.5) was the basis for Downward {\it et
al.}\cite{Downward2005} to propose the dimeron model.  However note that at any
point in time Mn sites occupied by a hole will be in the process of becoming
less distorted while the previously occupied site (which now has an e$_g$
electron) will be increasing in distortion.  Thus two sites are naturally
involved. If the hopping is slow enough that the e$_g$ sites become totally
distorted before being revisited by a hole then there should be much larger
distortions removed in the first 50\% of magnetization.  Alternatively if the
hopping is very fast more than two sites may have a low distortion. The small
decrease in distortion up to 50\% magnetization is thus evidence for fast
hopping on a local scale.

Alonso {\it et al.}\cite{Alonso00} also argue for localized Mn$^{+4}$ sites and
a reduce J-T distortion in the vicinity of Ca dopants, but do not provide a
reason for the reduced distortion. The EXAFS confirm that there is a low
distortion of many Mn sites - roughly two per Ca site. The number of Mn sites
in the vicinity of the Ca atoms is more than 4x (note that there are 8 Mn
neighbor sites per isolated Ca atom,  but only 4 additional Mn neighbor sites
for each Ca that is added to an existing linear chain of Ca) and thus the
number of low distortion sites must be more restricted than only counting the
number of closest Mn neighbors to Ca. If the hole can hop rapidly back and
forth between two sites, it explains the low distortion per site removed for
low magnetizations.  If the model of Alonso {\it et al.} is slightly modified
to correspond to partially delocalized holes in the vicinity of low J-T
distortions (i.e. a hole localized over $\sim$2 Mn sites) instead of a
completely localized Mn$^{+4}$ hole site, then their model becomes identical to
the dimeron model.

In summary, the new EXAFS data on LCMO samples with Ca concentrations near the
metal-insulator transition (x = 0.16-0.22) show similar behavior at the local
scale as do the CMR samples at higher Ca concentrations. The distortions
removed in the FM regime are small and show directly that there is a large
fraction of low distortion Mn sites associated with the charge carriers - the
dimeron sites.  A large fraction of the sample can be magnetized even though
the sample is insulating - which is quite similar to the CMR systems at
partial magnetization.  The unmagnetized fraction is presumably highly
distorted.


\acknowledgments 

This material is based upon work supported by the National Science Foundation.
The work at UCSC was supported by NSF grant DMR-0301971 and at Montana under
grant DMR-0504769.  The EXAFS experiments were performed at SSRL (operated by
the DOE, Division of Chemical Sciences, and by the NIH, Biomedical Resource
Technology Program, Division of Research Resources).

\newpage

\bibliographystyle{prsty}

\bibliography{/exafs/bib/bibli}

\end{document}